\documentclass[12pt]{article}
\usepackage{amsmath,amssymb,epsfig}
\makeatletter \@addtoreset{equation}{section} \makeatother
\renewcommand{\theequation}{\thesection.\arabic{equation}}
\newcommand{\ba}{\begin{array}}
\newcommand{\ea}{\end{array}}
\newcommand{\beq}{\begin{equation}}
\newcommand{\eeq}{\end{equation}}
\newcommand{\bea}{\begin{eqnarray}}
\newcommand{\eea}{\end{eqnarray}}

\def\bce{\begin{center}}
\def\ece{\end{center}}

\def\nonu{\nonumber}

\def\be{\beta}

\newcommand{\tr}{\mbox{Tr}}

\def\eps6{{\displaystyle \mathop{\epsilon}^{6}}{}}

\def\nab6{{\displaystyle \mathop{\nabla}^{6}}{}}

\def\0{{\sst{(0)}}}
\def\1{{\sst{(1)}}}
\def\2{{\sst{(2)}}}
\def\3{{\sst{(3)}}}
\def\4{{\sst{(4)}}}
\def\5{{\sst{(5)}}}
\def\6{{\sst{(6)}}}
\def\7{{\sst{(7)}}}
\def\8{{\sst{(8)}}}

\def\ba{\begin{array}}
\def\ea{\end{array}}
\def\beq{\begin{equation}}
\def\eeq{\end{equation}}
\def\be{\begin{equation}}
\def\ee{\end{equation}}

\def\tr{\mathop{\rm tr}}

\def\eps{\epsilon}

\def\ba{\begin{array}}
\def\ea{\end{array}}
\def\beq{\begin{equation}}
\def\eeq{\end{equation}}
\def\be{\begin{equation}}
\def\ee{\end{equation}}

\def\tr{\mathop{\rm tr}}

\def\eps{\epsilon}

\newcommand{\bean}{\begin{eqnarray*}}
\newcommand{\eean}{\end{eqnarray*}}

\begin{document}
\thispagestyle{empty} \addtocounter{page}{-1}
\begin{flushright}
\end{flushright}

\vspace*{1.3cm}

\centerline{ \Large \bf Meta-Stable Brane Configuration of Product
  Gauge Groups   }
\vspace*{1.5cm}
\centerline{{\bf Changhyun Ahn} 
} 
\vspace*{1.0cm} 
\centerline{\it 
Department of Physics, Kyungpook National University, Taegu
702-701, Korea} 
\vspace*{0.8cm} 
\centerline{\tt ahn@knu.ac.kr} 
\vskip2cm

\centerline{\bf Abstract}
\vspace*{0.5cm}
  
Starting from the ${\cal N}=1$  
$SU(N_c) \times SU(N_c')$ gauge theory with fundamental 
and bifundamental flavors, we apply the Seiberg dual 
to the first gauge group and obtain the ${\cal N}=1$
dual gauge theory with dual matters including the gauge
singlets. By analyzing the F-term equations of the  
superpotential, we describe the intersecting type IIA brane
configuration for the meta-stable nonsupersymmetric 
vacua of this gauge theory.
By introducing an orientifold 6-plane, we generalize to 
the case for ${\cal N}=1$  
$SU(N_c) \times SO(N_c')$ gauge theory with fundamental and  
bifundamental flavors.  
Finally, the ${\cal N}=1$  
$SU(N_c) \times Sp(N_c')$ gauge theory with matters 
is also described very briefly. 

\baselineskip=18pt
\newpage
\renewcommand{\theequation}
{\arabic{section}\mbox{.}\arabic{equation}}

\section{Introduction}

It is well-known that the ${\cal N}=1$ $SU(N_c)$ QCD with 
fundamental flavors
has a vanishing superpotential before we deform this theory by mass 
term for
quarks.
The vanishing
superpotential in the electric theory 
makes it easier to analyze its nonvanishing dual
magnetic superpotential.  
Sometimes by tuning the
various  rotation angles between NS5-branes and D6-branes
appropriately, even if the electric theory has nonvanishing superpotential,
one can make the nonzero
superpotential to vanish in the electric theory.  
Two procedures, deforming the electric gauge theory by adding the mass for the
quarks and taking the Seiberg dual magnetic theory from the electric 
theory, are
crucial to find out meta-stable supersymmetry breaking vacua in the
context of dynamical supersymmetry breaking \cite{ISS,IS}.
Some models of dynamical supersymmetry breaking can be studied by gauging the
subgroup of the
flavor symmetry group by either field theory analysis or using the brane
configuration \footnote{For the type IIA brane configuration 
description of ${\cal N}=1$ supersymmetric gauge theory, 
see the review paper \cite{GK}.}. 

In this paper, 
starting from the known ${\cal N}=1$ supersymmetric electric gauge 
theories, 
we construct the ${\cal
N}=1$ supersymmetric  magnetic
gauge theories by
brane motion and linking number counting. The dual gauge group appears
only on the first gauge group.
Based on their particular limits of corresponding 
magnetic brane configurations in the sense that their electric
theories do not have any superpotentials except the mass deformations for
the quarks, we 
describe the intersecting brane configurations of type IIA string
theory for the meta-stable nonsupersymmetric vacua 
of these gauge theories. 

We focus on the cases where 
the whole gauge group is given by a product of two gauge groups. 
One example 
can be realized by three NS5-branes with D4- and D6-branes, and 
the other by four NS5-branes with D4- and D6-branes.
For the latter, the appropriate orientifold 6-plane should be 
located  at the center of this brane configuration in order
to have two gauge groups. 
Of course, it is also possible, without changing the number of gauge
groups,  
to have the brane
configuration consisting of five NS5-branes and orientifold 6-plane,
at which the extra NS5-brane is located,  
with D4- and D6-branes, but we'll not do this particular 
case in this paper.     
 
In section 2, we review the type IIA brane configuration that contains
three NS5-branes, corresponding
to the electric theory based on the ${\cal N}=1$ $SU(N_c) \times
SU(N_c')$ 
gauge theory \cite{ILS,BH,BIWW}
with matter contents and deform this theory by adding the mass term
for the quarks. 
Then we construct the Seiberg dual magnetic theory which is 
${\cal N}=1$ $SU(\widetilde{N}_c) \times SU(N_c')$ gauge 
theory with corresponding dual
matters as well as various gauge singlets, by brane motion and linking
number counting. We do not touch the part of second gauge 
group $SU(N_c')$ in this dual process.

In section 3, we consider the nonsupersymmetric meta-stable
minimum by looking at the magnetic brane configuration we obtained in
section 2
and present 
the corresponding intersecting brane configuration of type IIA string
theory,  along the line of 
\cite{Ahn07-2,Ahn07-1,Ahn07,Ahn06-1,Ahn06}(see also \cite{OO,FGU,BGHSS}) and 
we describe M-theory lift of this supersymmetry breaking 
type IIA brane configuration.

In section 4, we describe the type IIA brane configuration that
contains four NS5-branes, corresponding
to the electric theory based on the ${\cal N}=1$ $SU(N_c) \times
SO(N_c')$ 
gauge theory \cite{LO}
with matter contents and deform this theory by adding the mass term
for the quarks. 
Then we take the Seiberg dual magnetic theory which is 
${\cal N}=1$ $SU(\widetilde{N}_c) \times SO(N_c')$ gauge 
theory with corresponding dual
matters as well as various gauge singlets, by brane motion and linking
number counting. The part of second gauge
group $SO(N_c')$ in this dual process is not changed under this process.

In section 5, the nonsupersymmetric meta-stable
minimum by looking at the magnetic brane configuration we obtained in
section 4 is constructed
and we present 
the corresponding intersecting brane configuration of type IIA string
theory and 
describe M-theory lift of this supersymmetry breaking 
type IIA brane configuration, as we did in section 3.

In section 6, we describe the similar application to the 
${\cal N}=1$ $SU(N_c) \times Sp(N_c')$ 
gauge theory \cite{LO} briefly 
and make some comments for the future directions.

\section{The ${\cal N}=1$ supersymmetric  
brane configuration  of $SU(N_c) \times SU(N_c')$ gauge theory}

After reviewing the type IIA brane configuration corresponding
to the electric theory based on the ${\cal N}=1$ $SU(N_c) \times
SU(N_c')$ 
gauge theory \cite{ILS,BH,BIWW}, 
we construct the Seiberg dual magnetic theory which is 
${\cal N}=1$ $SU(\widetilde{N}_c) \times SU(N_c')$ gauge 
theory.

\subsection{Electric theory with $SU(N_c) \times SU(N_c')$ gauge group}

The gauge group is given by $SU(N_c) \times SU(N_c')$ and the matter 
contents \cite{ILS,BH,BIWW} are given by 

$\bullet$
$N_f$ chiral multiplets $Q$ are 
in the fundamental representation under the $SU(N_c)$,
$N_f$ chiral multiplets $\widetilde{Q}$ are in the antifundamental
representation under the $SU(N_c)$ and then $Q$ are in the
representation $({\bf N_c,1
})$ while $\widetilde{Q}$ are in the representation $({\bf \overline{N_c}, 1})$
under the gauge group

$\bullet$
$N_f'$ chiral multiplets $Q'$ are 
in the fundamental representation under the $SU(N_c')$,
$N_f'$ chiral multiplets $\widetilde{Q'}$ are in the antifundamental
representation under the $SU(N_c')$ and then $Q'$ are in the
representation $({\bf 1, N_c'
})$ while $\widetilde{Q'}$ are in the representation $({\bf 1, \overline{N_c'}})$
under the gauge group

$\bullet$
The flavor singlet field $X$ is 
in the bifundamental representation $({\bf N_c, \overline{N_c'} })$ 
under the gauge group and its complex conjugate field $\widetilde{X}$
 is 
in the bifundamental representation $({\bf \overline{N_c}, N_c'})$ 
under the gauge group

In the electric theory, since there exist $N_f$ quarks $Q$, $N_f$
quarks $\widetilde{Q}$, one bifundamental field $X$ which will give
rise to the contribution of $N_c'$ and 
its complex conjugate $\widetilde{X}$ which will give
rise to the contribution of $N_c'$, the coefficient of the beta function
of the first gauge group factor is
\bea
b_{SU(N_c)} = 3N_c -N_f-N_c'
\nonu
\eea
and similarly
since there exist $N_f'$ quarks $Q'$, $N_f'$
quarks $\widetilde{Q'}$, one bifundamental field $X$ which will give
rise to the contribution of $N_c$ and 
its complex conjugate $\widetilde{X}$ which will give
rise to the contribution of $N_c$, the coefficient of the beta function
of the second gauge group factor is
\bea
b_{SU(N_c')} = 3N_c'-N_f'-N_c.
\nonu
\eea

The anomaly free global symmetry is given by $[SU(N_f) \times SU(N_f')]^2 \times
U(1)^3 \times U(1)_R$ \cite{ILS,BH,BIWW} and let us denote the
strong coupling scales for $SU(N_c)$ as $\Lambda_1$ and for $SU(N_c')$
as $\Lambda_2$.  
The theory is asymptotically free when $b_{SU(N_c)} = 3N_c -N_f-N_c' >
0$ for the $SU(N_c)$ gauge theory and when 
$b_{SU(N_c')} = 3N_c'-N_f'-N_c > 0$ for the $SU(N_c')$ gauge theory.

The type IIA brane configuration for this theory 
can be described by
$N_c$ color 
D4-branes (01236) suspended between a middle NS5-brane (012345) 
and the right NS5'-brane (012389) (denoted by $NS5_R'$-brane) along $x^6$
direction,
together with $N_f$ D6-branes (0123789) 
which are parallel to $NS5_R'$-brane and have nonzero (45) directions.
Moreover, an extra left NS5'-brane (denoted by $NS5_L'$-brane) 
is located at the left hand side of
a middle NS5-brane along the $x^6$ direction and there exist $N_c'$
color D4-branes
suspended 
between them, with  $N_f'$ D6-branes which have zero (45) directions. 
These are shown in Figure 1 explicitly. See also \cite{GK} for the
brane configuration.

By realizing that the two outer $NS5_{L,R}'$-branes are 
perpendicular to a middle NS5-brane and 
the fact that $N_f$ D6-branes are parallel to $NS5_R'$-brane
and $N_f'$ D6-branes are parallel to $NS5_L'$-brane,
the classical superpotential vanishes. However, one can deform this theory.
Then the classical superpotential by deforming this theory by adding the
mass term for the quarks $Q$ and $\widetilde{Q}$, along the lines of 
\cite{ISS,Ahn06,Ahn06-1,Ahn07,Ahn07-1,Ahn07-2}, is given by
\bea
W = m Q \widetilde{Q}
\label{superpotential}
\eea
and this type IIA brane configuration can be summarized as follows \footnote{We
introduce two complex coordinates $v \equiv x^4 + i x^5$ and $w \equiv
x^8 + i x^9$ for simplicity. }:

$\bullet$
One middle NS5-brane with worldvolume $(012345)$. 

$\bullet$ 
Two NS5'-branes
with worldvolume $(012389)$.  

$\bullet$
$N_f$ D6-branes with worldvolume $(0123789)$ located at
the 
positive region in $v$ direction. 

$\bullet$
$N_c$ color D4-branes with worldvolume $(01236)$. 
These are suspended between a middle 
NS5-brane and $NS5_R'$-brane.  

$\bullet$ 
$N_c'$ color D4-branes with worldvolume $(01236)$. These are suspended 
between $NS5_L'$-brane and a middle NS5-brane.

Now we draw this electric brane configuration in Figure 1 and we put
the coincident $N_f$ D6-branes in the nonzero $v$ direction.
If we ignore the left $NS5'_L$-brane, $N_c'$ D4-branes and $N_f'$
D6-branes, 
then this brane configuration 
corresponds to the standard ${\cal N}=1$ SQCD with the gauge group $SU(N_c)$ with
$N_f$ massive flavors.
The electric quarks $Q$ and $\widetilde{Q}$ correspond to strings
stretching between the
$N_c$ color D4-branes with $N_f$ D6-branes,
the electric quarks $Q'$ and $\widetilde{Q'}$ correspond to strings between the
$N_c'$ color D4-branes with $N_f'$ D6-branes and
the bifundamentals $X$ and $\widetilde{X}$ correspond to   strings 
stretching between the
$N_c$ color D4-branes with $N_c'$ color D4-branes. 

\begin{figure}[ht]
   \epsfxsize=4.0in 
\centerline{\epsffile{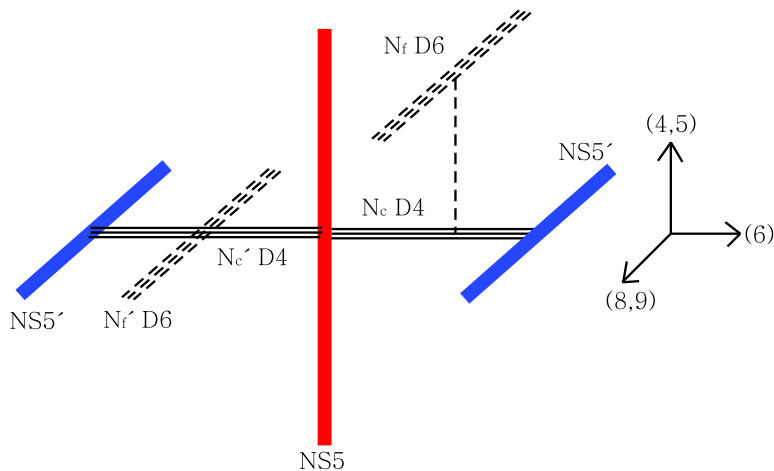}}
   \caption[FIG. \arabic{figure}.]{ 
The ${\cal N}=1$ supersymmetric electric brane configuration of 
$SU(N_c) \times SU(N_c')$ with $N_f$ chiral multiplets $Q$,
$N_f$ chiral multiplets $\widetilde{Q}$, $N_f'$ chiral multiplets $Q'$,
$N_f'$ chiral multiplets $\widetilde{Q'}$, the flavor singlet
bifundamental field $X$ and its 
complex conjugate bifundamental field $\widetilde{X}$. The $N_f$
D6-branes have nonzero $v$ coordinates where $v =m$ for equal massive
case of quarks $Q, \widetilde{Q}$ while $Q'$ and $\widetilde{Q'}$ are massless.   }
\label{fig1}
\end{figure}

\subsection{Magnetic theory with $SU(\widetilde{N}_c) \times SU(N_c')$
  gauge group}

One can consider dualizing one of the gauge groups regarding as the
other gauge group as a spectator. 
One takes the Seiberg dual for the first gauge group factor $SU(N_c)$
while remaining the second gauge group factor $SU(N_c')$ unchanged. 
Also we consider the case where $\Lambda_1 >> \Lambda_2$, in other
words, the dualized group's dynamical scale is far above that of the
other spectator group.

Let us move a middle NS5-brane to the right all the way past the right 
$NS5'_R$-brane. For example, see 
\cite{OO,FGU,BGHSS,Ahn06,Ahn06-1,Ahn07,Ahn07-1,Ahn07-2}. 
After this brane motion, one arrives at the Figure 2.
Note that there exists a creation of $N_f$ D4-branes
connecting $N_f$ D6-branes and $NS5'_R$-brane.
Recall that the $N_f$ D6-branes are perpendicular to a middle
NS5-brane in Figure 1.
The linking number \cite{HW} of NS5-brane from Figure 2
is 
$
L_5 = \frac{N_f}{2} -\widetilde{N}_c$.
On the other hand, the linking number of NS5-brane from Figure 1
is
$
L_5 = -\frac{N_f}{2} + N_c -N_c'$. Due to the connection of $N_c'$
D4-branes with $NS5_R'$-brane, the presence of $N_c'$ in the linking
number arises.
From these two relations, one obtains
the number of colors of dual magnetic theory
\bea
\widetilde{N}_c = N_f +N_c'-N_c.
\label{number}
\eea
The linking number counting looks similar to the one in \cite{Ahn07-2}
where there exists a contribution from O4-plane. 

Let us draw this magnetic brane configuration in Figure 2 and recall
that we put
the coincident $N_f$ D6-branes in the nonzero $v$ directions in the
electric theory, along the lines of 
\cite{OO,FGU,BGHSS,Ahn06,Ahn06-1,Ahn07,Ahn07-1,Ahn07-2}.
The $N_f$ created D4-branes connecting between
D6-branes and $NS5_R'$-brane can move freely in the $w$ direction.
Moreover since $N_c'$ D4-branes are suspending between two equal
$NS5'_{L,R}$-branes located at different $x^6$ coordinate, these D4-branes
can slide along the $w$ direction also.
If we ignore the left $NS5'_L$-brane, $N_c'$ D4-branes and $N_f'$ 
D6-branes(detaching these from Figure 2), 
then this brane configuration 
corresponds to the standard ${\cal N}=1$ SQCD with the magnetic gauge group 
$SU(\widetilde{N}_c=N_f-N_c)$ with
$N_f$ massive flavors \cite{OO,FGU,BGHSS}.

\begin{figure}[ht]
   \epsfxsize=4.0in 
\centerline{\epsffile{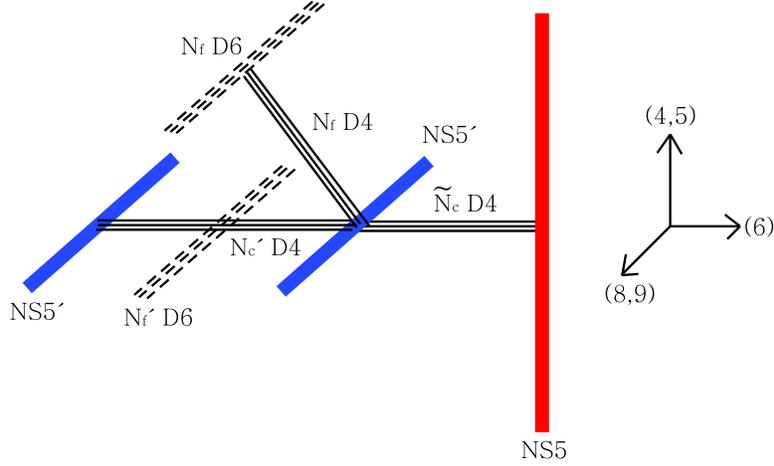}}
   \caption[FIG. \arabic{figure}.]{ 
The ${\cal N}=1$ supersymmetric magnetic brane configuration of 
$SU(\widetilde{N}_c=N_f +N_c'-N_c) \times SU(N_c')$ 
with $N_f$ chiral multiplets $q$,
$N_f$ chiral multiplets $\widetilde{q}$, $N_f'$ chiral multiplets $Q'$,
$N_f'$ chiral multiplets $\widetilde{Q'}$, the flavor singlet
bifundamental field $Y$ and its 
complex conjugate bifundamental field $\widetilde{Y}$
as well as
$N_f$ fields $F'$, its complex conjugate
$N_f$ fields $\widetilde{F'}$,
$N_f^2$ fields $M$ and the gauge singlet $\Phi$.
There exist $N_f$ flavor D4-branes connecting D6-branes and
$NS5_R'$-brane. }
\label{fig2}
\end{figure}

The dual magnetic gauge group
is given by $SU(\widetilde{N}_c) \times SU(N_c')$ and the matter
contents are given by

$\bullet$ 
$N_f$ chiral multiplets $q$ are 
in the fundamental representation under the $SU(\widetilde{N}_c)$,
$N_f$ chiral multiplets $\widetilde{q}$ are in the antifundamental
representation under the $SU(\widetilde{N}_c)$ and then $q$ are in the
representation $({\bf \widetilde{N}_c,1
})$ while $\widetilde{q}$ are in the representation 
$({\bf \overline{\widetilde{N}_c}, 1})$
under the gauge group

$\bullet$
$N_f'$ chiral multiplets $Q'$ are 
in the fundamental representation under the $SU(N_c')$,
$N_f'$ chiral multiplets $\widetilde{Q'}$ are in the antifundamental
representation under the $SU(N_c')$ and then $Q'$ are in the
representation $({\bf 1, N_c'
})$ while $\widetilde{Q'}$ are in the representation $({\bf 1, \overline{N_c'}})$
under the gauge group

$\bullet$
The flavor singlet field $Y$ is 
in the bifundamental representation $({\bf \widetilde{N}_c, \overline{N_c'} })$ 
under the gauge group and its complex conjugate field $\widetilde{Y}$
 is 
in the bifundamental representation $({\bf \overline{\widetilde{N}_c}, N_c'})$ 
under the gauge group

There are $(N_f+N_c')^2$ gauge singlets in the first dual gauge group
factor
as follows:

$\bullet$
$N_f$-fields $F'$ are in the fundamental representation under the
$SU(N_c')$, its complex conjugate
$N_f$-fields $\widetilde{F'}$ are in the antifundamental
representation under the $SU(N_c')$ and then 
$F'$ are in the representation $({\bf 1, N_c' })$ 
under the gauge group
while 
$\widetilde{F'}$ are in the representation $({\bf 1, \overline{N_c'} })$ 
under the gauge group

These additional $N_f$ $SU(N_c')$ fundamentals and $N_f$ $SU(N_c')$
antifundamentals 
are originating from 
the $SU(N_c)$ chiral mesons $\widetilde{X} Q$ and $X \widetilde{Q}$ respectively.
It is clear to see that from the Figure 2, since the
$N_f$ D6-branes are parallel to the $NS5'_R$-brane, 
the newly created $N_f$ D4-branes can slide along the plane
consisting of these $D6$-branes and  $NS5'_R$-brane
arbitrarily.
Then strings stretching between the $N_f$ $D6$-branes and $N_c'$
D4-branes will give rise to these additional $N_f$ $SU(N_c')$
fundamentals and  $N_f$ $SU(N_c')$
antifundamentals.

$\bullet$
$N_f^2$-fields $M$ are in the representation $({\bf 1, 1})$ under the
gauge group

This corresponds to the $SU(N_c)$ chiral meson $Q \widetilde{Q}$ and
the fluctuations of the singlet $M$ correspond to the motion of $N_f$
flavor D4-branes along
(789) directions in Figure 2.

$\bullet$
The $N_c'^2$-fields 
$\Phi$ is in the representation $({\bf 1, N_c'^2-1}) \oplus ({\bf 1,1
})$ 
under the gauge group  

This corresponds to the $SU(N_c)$ chiral meson $X \widetilde{X}$ and
note that  
$X$  has a representation ${\bf \overline{N_c'}}$ of $SU(N_c')$ while
$\widetilde{X}$
has a representation ${\bf N_c'}$ of $SU(N_c')$.
The fluctuations of the singlet $\Phi$ correspond to the motion of
$N_c'$ D4-branes suspended two $NS5_{L,R}'$-branes along the (789)
directions in Figure 2. 

In the dual theory, 
since there exist $N_f$ quarks $q$, $N_f$
quarks $\widetilde{q}$, one bifundamental field $Y$ which will give
rise to the contribution of $N_c'$ and 
its complex conjugate $\widetilde{Y}$ which will give
rise to the contribution of $N_c'$,
the coefficient of the beta function for the first gauge group factor \cite{BIWW}
is
\bea
b_{SU(\widetilde{N}_c)}^{mag}
= 3\widetilde{N}_c-N_f-N_c'
=2N_f +2N_c'-3N_c
\nonu
\eea
where we inserted the number of colors given in (\ref{number}) in the
second equality
and 
since there exist $N_f'$ quarks $Q'$, $N_f'$
quarks $\widetilde{Q'}$, one bifundamental field $Y$ which will give
rise to the contribution of $\widetilde{N}_c$, 
its complex conjugate $\widetilde{Y}$ which will give
rise to the contribution of $\widetilde{N}_c$,
$N_f$ fields $F'$, its complex conjugate
$N_f$ fields $\widetilde{F'}$ and the singlet $\Phi$ which will give
rise to $N_c'$, 
the coefficient of the beta function of second gauge group factor \cite{BIWW}
 is
\bea
b_{SU(N_c')}^{mag} 
= 3N_c'-N_f'-\widetilde{N}_c-N_f-N_c' 
=N_c'+N_c-2N_f-N_f'.
\nonu
\eea
Therefore, both $SU(\widetilde{N}_c)$ and 
$SU(N_c')$ gauge couplings are IR free
by requiring the negativeness of the coefficients of beta function.
One can rely on the perturbative calculations at low energy 
for this magnetic IR free region $b_{SU(\widetilde{N}_c)}^{mag} < 0$ and 
$b_{SU(N_c')}^{mag} < 0$.
Note that the $SU(N_c')$ fields in the magnetic theory 
are different from those of the electric theory.
Since $b_{SU(N_c')}-b_{SU(N_c')}^{mag} > 0$, $SU(N_c')$ is more
asymptotically free than $SU(N_c')^{mag}$ \cite{BIWW}.
Neglecting the $SU(N_c')$ dynamics, the magnetic $SU(\widetilde{N}_c)$
is IR free when $N_f+N_c'< \frac{3}{2} N_c$ \cite{BIWW}.

The dual magnetic superpotential, by adding the mass term 
(\ref{superpotential}) for $Q$ and
$\widetilde{Q}$
in the electric theory which is equal to put a linear term in $M$ in
the dual magnetic theory, is given by 
\bea
W_{dual} = \left(M q \widetilde{q} + Y F' \widetilde{q} + \widetilde{Y} q
\widetilde{F'} + \Phi Y \widetilde{Y} \right) + m M
\label{dualW}
\eea
where the mesons in terms of the fields defined in the electric theory
are
\bea
 M \equiv Q \widetilde{Q}, \qquad \Phi \equiv X \widetilde{X}, \qquad 
F' \equiv \widetilde{X} Q, \qquad \widetilde{F'} \equiv X \widetilde{Q}.
\nonu 
\eea
By orientifolding procedure(O4-plane) into this brane configuration,
the $q(Q)$ and $\widetilde{q}(\widetilde{Q})$ are equivalent to each other, 
the $Y(X)$ and $\widetilde{Y}(\widetilde{X})$ become identical and 
$F'$ and $\widetilde{F'}$ become the same.
Then the reduced superpotential is identical with the one in \cite{Ahn07-2}. 
Here $q$ and $\widetilde{q}$ are fundamental and antifundamental for
the gauge group index respectively and antifundamentals for the flavor index.
Then, $q \widetilde{q}$ has rank $\widetilde{N}_c$ while $m$ has a
rank $N_f$.  Therefore, the F-term condition, the derivative the 
superpotential $W_{dual}$ with respect to $M$, cannot be satisfied 
if the rank $N_f$ exceeds $\widetilde{N}_c$. 
This is so-called rank condition and the supersymmetry is broken.    
Other F-term equations are satisfied by taking the vacuum expectation 
values of $Y, \widetilde{Y}, F'$ and $\widetilde{F'}$ to vanish.

The classical moduli space of vacua can be obtained from F-term
equations
\bea
 q  \widetilde{q} +  m & = & 0, \qquad
\widetilde{q} M + \widetilde{F'} \widetilde{Y}   =  0, \nonu \\
  M q  + Y F' & = & 0, \qquad
F' \widetilde{q} +  \widetilde{Y} \Phi  =  0,
\nonu \\
\widetilde{q} Y & = & 0, \qquad
q \widetilde{F'} +  \Phi Y  =  0, \nonu \\
\widetilde{Y} q & = & 0, \qquad
Y \widetilde{Y}  =  0. 
\nonu
\eea
Then, it is easy to see that there exist three reduced  equations
\bea
\widetilde{q} M =0= M q, \qquad
 q  \widetilde{q} +  m  =  0
\nonu 
\eea
and other F-term equations are satisfied if one takes the zero vacuum
expectation values for the fields $Y, \widetilde{Y}, F'$ and 
$\widetilde{F'}$.
Then the solutions can be written as follows:
\bea
<q >  & = &  \left(
\begin{array}{c}
\sqrt{m} e^{\phi} {\bf 1}_{\widetilde{N}_c}  \\
0
\end{array}
\right),  
< \widetilde{q}> =
 \left(
\begin{array}{cc}
\sqrt{m} e^{-\phi}  {\bf 1}_{\widetilde{N}_c}   &
0
\end{array}
\right), 
<M>  =
 \left(
\begin{array}{cc}
0  & 0 
 \\
0 & \Phi_0  {\bf 1}_{N_f-\widetilde{N}_c} 
\end{array}
\right)
\nonu \\
<Y> & = & <\widetilde{Y}> = <F'> = <\widetilde{F'}>= 0.
\label{vac}
\eea
Let us expand around  a point on (\ref{vac}), as done in
\cite{ISS}. 
Then the remaining relevant terms of superpotential are given by
\bea
W_{dual}^{rel} & = &  \Phi_0 \left( \delta \varphi  
\; \delta \widetilde{\varphi} + m \right) +
  \delta Z \; \delta \varphi  \; \widetilde{q}_0 
+ \delta \widetilde{Z} \; q_0  
\delta \widetilde{\varphi}
\nonu
\eea
by following the same 
fluctuations for the various fields as in \cite{Ahn07}:
\bea
q  & = &
\left(
\begin{array}{c}
q_0  {\bf 1}_{\widetilde{N}_c} +\frac{1}{\sqrt{2}}(\delta \chi_{+} + \delta \chi_{-})
 {\bf 1}_{\widetilde{N}_c} \nonu \\
\delta \varphi
\end{array}
\right), 
\quad 
\widetilde{q}   = 
 \left(
\begin{array}{cc}
\widetilde{q}_0   {\bf 1}_{\widetilde{N}_c} +
\frac{1}{\sqrt{2}}(\delta \chi_{+} - \delta \chi_{-})
  {\bf 1}_{\widetilde{N}_c}   &
\delta \widetilde{\varphi}
\end{array}
\right),
\nonu \\
M  & = &
 \left(
\begin{array}{cc}
\delta Y  & \delta Z
 \\
\delta \widetilde{Z} & \Phi_0  {\bf 1}_{N_f-\widetilde{N}_c} 
\end{array}
\right)
\nonu
\eea
as well as the fluctuations of $Y, \widetilde{Y}, F'$ and 
$\widetilde{F'}$.
Note that there exist also three kinds of terms, 
the vacuum  $<q>$ multiplied by 
$\delta \widetilde{Y} \delta \widetilde{F'}$,  
the vacuum  $<\widetilde{q}>$ multiplied by $\delta F' 
\delta Y$, and 
the vacuum  $<\Phi>$ multiplied by $\delta Y 
\delta \widetilde{Y}$. However,
by redefining these, they do not enter the 
contributions for the one loop result, up to quadratic order. 
As done in \cite{Shih}, one gets 
that $m_{\Phi_0}^2$ will contain $(\log 4 -1) > 0$ implying that these
are stable.

\section{Nonsupersymmetric meta-stable brane configuration
of $SU(N_c) \times SU(N_c')$ gauge theory }

Now we recombine $\widetilde{N}_c$ D4-branes among $N_f$ flavor D4-branes 
connecting between D6-branes and $NS5_R'$-brane with those 
connecting between $NS5'_R$-brane and NS5-brane and push
them in $+v$ direction from Figure 2. 
After this procedure, there are no color D4-branes between 
$NS5'_R$-brane and NS5-brane.
For the flavor D4-branes, we are left with only 
$(N_f-\widetilde{N}_c)$ flavor D4-branes.  

Then the minimal energy supersymmetry breaking brane configuration is
shown in Figure 3, along the lines of 
\cite{OO,FGU,BGHSS,Ahn06,Ahn06-1,Ahn07,Ahn07-1,Ahn07-2}.
If we ignore the left $NS5'_L$-brane, $N_c'$ D4-branes and 
$N_f'$ D6-branes(detaching these from Figure 3), 
as observed already, 
then this brane configuration 
corresponds to  the minimal energy supersymmetry breaking brane
configuration
for the ${\cal N}=1$ SQCD with the magnetic gauge group 
$SU(\widetilde{N}_c=N_f-N_c)$ with
$N_f$ massive flavors \cite{OO,FGU,BGHSS}.

\begin{figure}[ht]
   \epsfxsize=4.0in 
\centerline{\epsffile{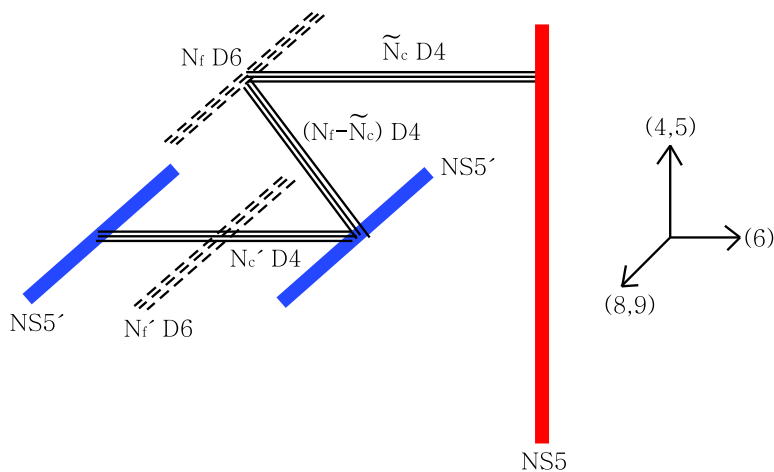}}
   \caption[FIG. \arabic{figure}.]{ 
The nonsupersymmetric minimal energy brane configuration of 
$SU(\widetilde{N}_c=N_f +N_c'-N_c) \times SU(N_c')$ 
with $N_f$ chiral multiplets $q$,
$N_f$ chiral multiplets $\widetilde{q}$, $N_f'$ chiral multiplets $Q'$,
$N_f'$ chiral multiplets $\widetilde{Q'}$, the flavor singlet
bifundamental field $Y$ and its 
complex conjugate bifundamental field $\widetilde{Y}$ and 
various gauge singlets.  }
\label{fig3}
\end{figure}

The type IIA/M-theory brane construction for the ${\cal N}=2$
gauge theory was described by \cite{Witten} and after lifting 
the type IIA description to M-theory, the 
corresponding magnetic M5-brane configuration \footnote{The M5-brane
  lives in (0123) directions and is wrapping on a Riemann surface
  inside (4568910) directions. The Taub-NUT space in (45610)
  directions is parametrized by two complex variables $v$ and $y$ and
  the flat two dimensions in (89) directions by a complex variable
  $w$.
See \cite{BGHSS} for the relevant discussions. }
with equal mass for the
quarks where the gauge group is given by 
$SU(\widetilde{N}_c) \times SU(N_c')$, in a background space of $x t = 
v^{N_f'} \prod_{k=1}^{N_f} (v -e_k)$ where this four dimensional space
replaces (45610) directions,
is described by 
\bea
t^3 + ( v^{\widetilde{N}_c }  + \cdots ) t^2 + 
( v^{N_c'} + \cdots) t +  v^{N_f'}
\prod_{k=1}^{N_f} (v -e_k ) =0
\label{curve}
\eea
where $e_k$ is the position of the D6-branes in the $v$ direction(for
equal massive case, we can write $e_k=m$) and
we have ignored the lower power terms in $v$ in $t^2$ and $t$ 
denoted by $\cdots$
and the scales for the gauge groups in front of the first term and the
last term, for simplicity. For fixed $x$, the coordinate $t$
corresponds to $y$.

From this curve (\ref{curve}) 
of cubic equation for $t$ above, the asymptotic regions 
for three NS5-branes can be classified by looking at 
the first two terms providing NS5-brane asymptotic region, 
next two terms providing $NS5_R'$-brane asymptotic region and 
the final two terms giving $NS5_L'$-brane asymptotic region
as follows

1. $v \rightarrow \infty$ limit implies
\bea
w \rightarrow 0, \quad y \sim    v^{\widetilde{N}_c} + \cdots \quad
\mbox{NS asymptotic region}.   
\nonu
\eea

2.  $w \rightarrow \infty$ limit implies
\bea
v & \rightarrow &   m, \quad 
y \sim  w^{N_f+N_f'-N_c'}
 +\cdots
\quad \mbox{$NS_{L}'$ asymptotic region}, 
\nonu
\\
v & \rightarrow &  m, \quad  
y \sim w^{N_c'-\widetilde{N}_c}
+\cdots
\quad \mbox{$NS_{R}'$ asymptotic region}. 
\nonu
\eea
Here the two $NS5'_{L,R}$-branes are moving in the
$+v$ direction holding everything else fixed instead of moving
D6-branes in the $+v$ direction, in the spirit of \cite{BGHSS}. 
The harmonic function sourced by the D6-branes can be written
explicitly by summing over two contributions from the 
$N_f$ and $N_f'$ D6-branes and similar analysis to both solve
the differential equation and find out the nonholomorphic curve can be
done
\cite{BGHSS,Ahn06-1,Ahn07,Ahn07-1,Ahn07-2}. 
An instability from a new M5-brane mode arises.

\section{The ${\cal N}=1$ supersymmetric 
brane configuration of $SU(N_c) \times SO(N_c')$ gauge theory }

After reviewing the type IIA brane configuration corresponding
to the electric theory based on the ${\cal N}=1$ $SU(N_c) \times
SO(N_c')$ 
gauge theory \cite{LO}, 
we describe the Seiberg dual magnetic theory which is 
${\cal N}=1$ $SU(\widetilde{N}_c) \times SO(N_c')$ gauge 
theory.

\subsection{Electric theory with $SU(N_c) \times SO(N_c')$ gauge group }

The gauge group is given by $SU(N_c) \times SO(N_c')$ and the matter 
contents \cite{LO}(similar matter contents are found in \cite{ILS}) are given by

$\bullet$
$N_f$ chiral multiplets $Q$ are 
in the fundamental representation under the $SU(N_c)$,
$N_f$ chiral multiplets $\widetilde{Q}$ are in the antifundamental
representation under the $SU(N_c)$ and then $Q$ are in the
representation $({\bf N_c,1
})$ while $\widetilde{Q}$ are in the representation $({\bf \overline{N_c}, 1})$
under the gauge group

$\bullet$
$2N_f'$ chiral multiplets $Q'$ are 
in the fundamental representation under the $SO(N_c')$ and then $Q'$ are in the
representation $({\bf 1, N_c'
})$
under the gauge group

$\bullet$
The flavor singlet field $X$ is 
in the bifundamental representation $({\bf N_c, N_c' })$ 
under the gauge group and the flavor singlet $\widetilde{X}$
 is 
in the bifundamental representation $({\bf \overline{N_c}, N_c'})$ 
under the gauge group

In the electric theory, since there exist $N_f$ quarks $Q$, $N_f$
quarks $\widetilde{Q}$, one bifundamental field $X$ which will give
rise to the contribution of $N_c'$ and 
its complex conjugate $\widetilde{X}$ which will give
rise to the contribution of $N_c'$, the coefficient of the beta
function
of the first gauge group factor
is
\bea
b_{SU(N_c)} = 3N_c -N_f-N_c'
\nonu
\eea
and similarly,
since there exist $2N_f'$ quarks $Q'$, one bifundamental field $X$ which will give
rise to the contribution of $N_c$ and 
its complex conjugate $\widetilde{X}$ which will give
rise to the contribution of $N_c$, the coefficient of the beta function
of the second gauge group factor is
\bea
b_{SO(N_c')} = 3(N_c'-2)-2N_f'-2N_c.
\nonu
\eea

The anomaly free global symmetry is given by $SU(N_f)^2 \times SU(2N_f') \times
U(1)^2 \times U(1)_R$ and let us denote the
strong coupling scales for $SU(N_c)$ as $\Lambda_1$ and for $SO(N_c')$
as $\Lambda_2$, as in previous section.  
The theory is asymptotically free when $b_{SU(N_c)}  >
0$ for the $SU(N_c)$ gauge theory and when 
$b_{SO(N_c')}  > 0$ for the $SO(N_c')$ gauge theory.

The type IIA 
brane configuration of ${\cal N}=2$ gauge theory 
\cite{LL} consists of four NS5-branes (012345)
which have different $x^6$ values, $N_c$ and $N_c'$
D4-branes (01236)
suspended between them, $2N_f$ and $2N_f'$ D6-branes (0123789) and an
orientifold 6 plane (0123789) of positive Ramond charge
\footnote{There are many different brane configurations with O6-plane
  in the literature and some of them are present 
in \cite{LLL,AOT,LLL1,BHKL,EGKT}.
}. 
According to ${\bf Z}_2$ symmetry of orientifold
6-plane(O6-plane) 
sitting at $v=0$ and $x^6=0$,
the coordinates $(x^4,x^5, x^6)$ transform as $-(x^4, x^5, x^6)$, as usual. 
See also \cite{GK} for the discussion of O6-plane.

By rotating the third and fourth NS5-branes which are located at the
right hand side of O6-plane, from $v$ direction toward
$-w$ and $+w$ 
directions respectively, 
one obtains 
${\cal N}=1$ theory.  
Their mirrors, the first and second NS5-branes which are located at the left hand
side of O6-plane, can be rotated in a
${\bf Z}_2$ symmetric manner due to the presence of O6-plane simultaneously. 
That is, if the first NS5-brane rotates by an angle $-\omega$ in
$(v,w)$ plane, denoted by  $NS5_{-\omega}$-brane \cite{GK}, then 
the mirror image of
this NS5-brane, the fourth NS5-brane, is rotated by an angle $\omega$
in the same plane, denoted by 
$NS5_{\omega}$-brane.  
If the second NS5-brane rotates by an angle $\theta$ in
$(v,w)$ plane, denoted by  $NS5_{\theta}$-brane \cite{GK}, then 
the mirror image of
this NS5-brane, the third NS5-brane, is rotated by an angle $-\theta$
in the same plane, denoted by 
$NS5_{-\theta}$-brane. 
For more details, see the Figure 4 \footnote{The angles of $\theta_1$
  and $\theta_2$ in \cite{LO} are related to the angles $\theta$ and
  $\omega$
as follows: $\theta= \theta_1$ and 
$\omega=\theta_2$.}.

We also rotate the $N_f'$ D6-branes which are located between the
second NS5-brane and an O6-plane and make them be parallel to
$NS5_{\theta}$-brane and denote them as $D6_{\theta}$-brane with zero
$v$ coordinate(the
angle between the unrotated D6-branes and $D6_{\theta}$-branes is
equal to $\frac{\pi}{2}-\theta$) and its mirrors $N_f'$ D6-branes appear
as $D6_{-\theta}$-branes between the O6-plane and third NS5-brane. 
There is no coupling between the adjoint field and
the quarks since the rotated $D6_{\theta}$-branes are parallel to the rotated
$NS5_{\theta}$-brane \cite{BH,GK}.
Similarly, the $N_f$ D6-branes which are located between the
third NS5-brane and the fourth NS5-brane can be rotated and we can 
make them be parallel to
$NS5_{\omega}$-brane and denote them as $D6_{\omega}$-branes with
nonzero $v$ coordinate(the
angle between the unrotated D6-branes and $D6_{\omega}$-branes is
equal to $\frac{\pi}{2}-\omega$) and its mirrors $N_f$ D6-branes appear
as $D6_{-\omega}$-branes between the first NS5-brane and the second NS5-brane. 

Moreover the $N_c$ D4-branes are suspended between the first NS5-brane
and the second NS5-brane(and its mirrors) and the $N_c'$ D4-branes are
suspended
between the second NS5-brane and the third NS5-brane.

For this brane setup \footnote{For arbitrary angles $\theta$ and
  $\omega$, the superpotential for the $SU(N_c)$ sector is given by
  $W=X \phi \widetilde{X} + \tan (\omega-\theta) \tr \phi^2$ where
  $\phi$ ia an adjoint field for $SU(N_c)$. There is no coupling
  between $\phi$ and $N_f$ quarks because $D6_{\pm \omega}$-branes are parallel
  to $NS5_{\pm \omega}$-branes. The superpotential for the $SO(N_c')$
sector is given by $W=X \phi_A \widetilde{X} + X \phi_S \widetilde{X}+
\tan \theta \tr \phi_A^2 -\frac{1}{\tan \theta} \tr \phi_S^2$ where 
$\phi_A$ and $\phi_S$ are an adjoint field and a symmetric tensor for
  $SO(N_c')$ \cite{CSST}.
After integrating out $\phi, \phi_A$ and $\phi_S$, the whole
  superpotential can be written as in (\ref{sup}).}, 
the classical superpotential is
given by \cite{LO}
\bea
W= -\frac{1}{4} \left[ \frac{1}{4 \tan(\omega - \theta)} +
  \frac{1}{\tan 2\theta} \right]
\tr (X \widetilde{X})^2 + \frac{ \tr X \widetilde{X}
\widetilde{X} X }{4 \sin 2\theta }
+ \frac{(\tr
X \widetilde{X})^2}{4 N_c \tan(\omega-\theta)}.
\label{sup}
\eea
It is easy to see that when $\theta$ approaches $0$ and
$\omega$
approaches $\frac{\pi}{2}$, 
then this superpotential vanishes.  

Now one summarizes
the supersymmetric electric brane configuration with their
worldvolumes in type IIA string theory as follows.

$\bullet$
$NS5_{-\omega}$-brane with worldvolume by both (0123) and two spatial dimensions
in $(v,w)$ plane and with negative $x^6$.

$\bullet$
$NS5_{\theta}$-brane with worldvolume by both (0123) and two spatial dimensions
in $(v,w)$ plane and with negative $x^6$.

$\bullet$
 $NS5_{-\theta}$-brane  with worldvolume by both (0123) and two spatial dimensions
in $(v,w)$ plane and with positive $x^6$.

$\bullet$
$NS5_{\omega}$-brane with worldvolume by both (0123) and two spatial dimensions
in $(v,w)$ plane and with positive $x^6$.

$\bullet$  $N_f'$
 $D6_{\theta}$-branes  with worldvolume by both (01237) and two spatial dimensions
in $(v,w)$ plane and with negative $x^6$ and $v=0$. 

$\bullet$  $N_f'$
$D6_{-\theta}$-branes  with worldvolume by both (01237) and two space dimensions
in $(v,w)$ plane and with positive $x^6$ and $v=0$.

$\bullet$  $N_f$
 $D6_{\omega}$-branes  with worldvolume by both (01237) and two spatial dimensions
in $(v,w)$ plane and with positive $x^6$. Before the rotation, the
 distance from $N_c$ color D4-branes in the $+v$ direction is nonzero.

$\bullet$  $N_f$
$D6_{-\omega}$-branes  with worldvolume by both (01237) and two space dimensions
in $(v,w)$ plane and with negative $x^6$.
 Before the rotation, the
 distance from $N_c$ color D4-branes in the $-v$ direction is nonzero.

$\bullet$ O6-plane  with worldvolume (0123789) with $v=0=x^6$.

$\bullet$ $N_c$ D4-branes connecting $NS5_{-\omega}$-brane and
$NS5_{\theta}$-brane,  
with worldvolume (01236) with $v=0=w$(and its mirrors).

$\bullet$ $N_c'$ D4-branes  connecting $NS5_{\theta}$-brane and
$NS5_{-\theta}$-brane,   with worldvolume (01236) with $v=0=w$.

We draw the type IIA electric brane configuration in Figure 4 which 
was basically given in
\cite{LO} already but the only difference is to put $N_f$ D6-branes in the
nonzero $v$ direction in order to obtain nonzero masses for the quarks
which are necessary to obtain the meta-stable vacua.

\begin{figure}[ht]
   \epsfxsize=5.0in 
\centerline{\epsffile{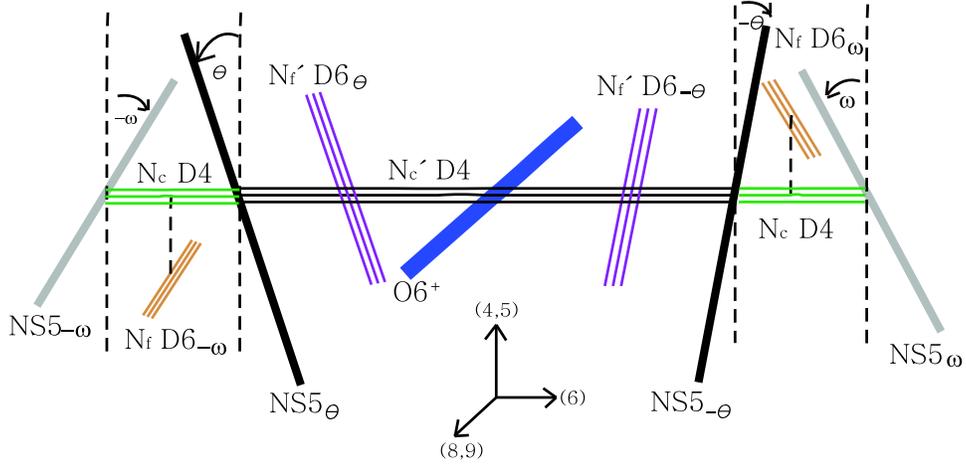}}
   \caption[FIG. \arabic{figure}.]{ 
The ${\cal N}=1$ supersymmetric electric brane configuration of 
$SU(N_c) \times SO(N_c')$ with $N_f$ chiral multiplets $Q$,
$N_f$ chiral multiplets $\widetilde{Q}$, $2N_f'$ chiral multiplets
$Q'$, 
the flavor singlet
bifundamental field $X$ and its 
complex conjugate bifundamental field $\widetilde{X}$. The $N_f$
$D6_{\omega}$-branes 
have nonzero $v$ coordinates where $v =  m$(and its mirrors) for equal massive
case of quarks $Q, \widetilde{Q}$ while $Q'$ is  massless. }
\label{fig4}
\end{figure}

\subsection{Magnetic theory with $SU(\widetilde{N}_c) \times SO(N_c')$
gauge group}

One takes the Seiberg dual for the first gauge group factor $SU(N_c)$
while remaining the second gauge group factor $SO(N_c')$, as in
previous case. 
Also we consider the case where $\Lambda_1 >> \Lambda_2$, in other
words, the dualized group's dynamical scale is far above that of the
other spectator group.

Let us move the $NS5_{-\theta}$-brane to the right all the way past the right 
$NS5_{\omega}$-brane(and its mirrors to the left).  
After this brane motion, one arrives at the Figure 5.
Note that there exists a creation of $N_f$ D4-branes
connecting $N_f$ $D6_{\omega}$-branes and $NS5_{\omega}$-brane(and its mirrors).
Recall that the $N_f$ $D6_{\omega}$-branes are not parallel to the
$NS5_{-\theta}$-brane in Figure 4(and its mirrors).
The linking number of $NS5_{-\theta}$-brane from Figure 5
is 
$
L_5 = \frac{N_f}{2} -\widetilde{N}_c$.
On the other hand, the linking number of $NS5_{-\theta}$-brane from Figure 4
is
$
L_5 = -\frac{N_f}{2} + N_c -N_c'$.
From these, one gets the number of colors in dual magnetic theory
\bea
\widetilde{N}_c = N_f +N_c'-N_c.
\label{num}
\eea

Let us draw this magnetic brane configuration in Figure 5 and remember
that we put
the coincident $N_f$ $D6_{\omega}$-branes in the nonzero $v$
directions(and its mirrors).
The $N_f$ created D4-branes connecting between
$D6_{\omega}$-branes and $NS5_{\omega}$-brane 
can move freely in the $w$ direction, as in previous case.
Moreover, since $N_c'$ D4-branes are suspending between two unequal
$NS5_{\pm \omega}$-branes located at different $x^6$ coordinate, these D4-branes
cannot slide along the $w$ direction, for arbitrary rotation angles.
If we are detaching all the branes except $NS5_{\omega}$-brane,
$NS5_{-\theta}$-brane, $D6_{\omega}$-branes, $N_f$ D4-branes and $\widetilde{N}_c$
D4-branes 
from Figure 5, 
then this brane configuration 
corresponds to ${\cal N}=1$ SQCD with the magnetic gauge group 
$SU(\widetilde{N}_c=N_f-N_c)$ with
$N_f$ massive flavors with tilted NS5-branes.

\begin{figure}[ht]
   \epsfxsize=5in 
\centerline{\epsffile{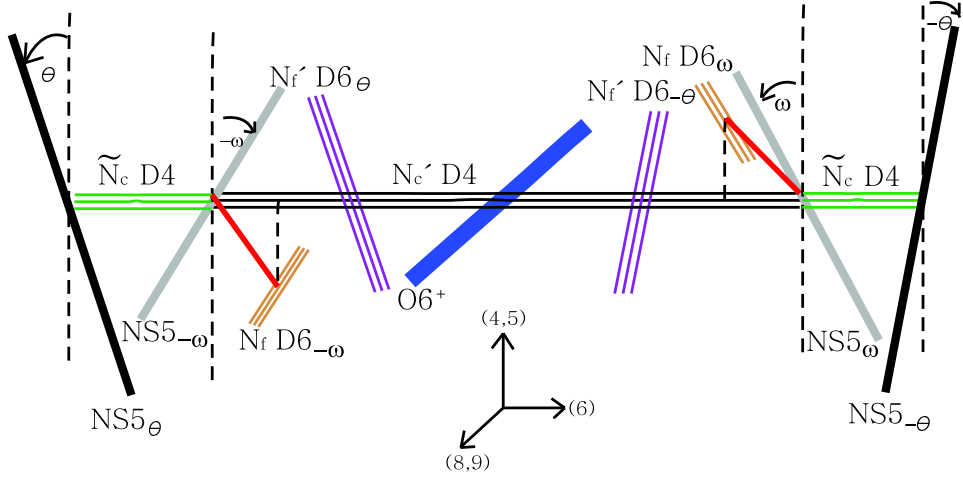}}
   \caption[FIG. \arabic{figure}.]{ 
The ${\cal N}=1$ supersymmetric magnetic brane configuration of 
$SU(\widetilde{N}_c=N_f +N_c'-N_c) \times SO(N_c')$ 
with $N_f$ chiral multiplets $q$,
$N_f$ chiral multiplets $\widetilde{q}$, $2N_f'$ chiral multiplets
$Q'$, 
the flavor singlet
bifundamental field $Y$ and its 
complex conjugate bifundamental field $\widetilde{Y}$
as well as
$N_f$ fields $F'$, its complex conjugate
$N_f$ fields $\widetilde{F'}$,
$N_f^2$ fields $M$ and the gauge singlet $\Phi$.
There exist $N_f$ flavor D4-branes connecting $D6_{\omega}$-branes and
$NS5_{\omega}$-brane(and its mirrors).}
\label{fig5}
\end{figure}

The dual magnetic gauge group
is given by $SU(\widetilde{N}_c) \times SO(N_c')$ and the matter
contents are given by

$\bullet$ 
$N_f$ chiral multiplets $q$ are 
in the fundamental representation under the $SU(\widetilde{N}_c)$,
$N_f$ chiral multiplets $\widetilde{q}$ are in the antifundamental
representation under the $SU(\widetilde{N}_c)$ and then $q$ are in the
representation $({\bf \widetilde{N}_c,1
})$ while $\widetilde{q}$ are in the representation 
$({\bf \overline{\widetilde{N}_c}, 1})$
under the gauge group

$\bullet$
$2N_f'$ chiral multiplets $Q'$ are 
in the fundamental representation under the $SO(N_c')$ and then $Q'$ are in the
representation $({\bf 1, N_c'
})$
under the gauge group

$\bullet$
The flavor singlet field $Y$ is 
in the bifundamental representation $({\bf \widetilde{N}_c, N_c' })$ 
under the gauge group and its complex conjugate field $\widetilde{Y}$
 is 
in the bifundamental representation $({\bf \overline{\widetilde{N}_c}, N_c'})$ 
under the gauge group

There are $(N_f+N_c')^2$ 
gauge singlets in the first dual gauge group factor as follows:

$\bullet$
$N_f$-fields $F'$ are in the fundamental representation under the
$SO(N_c')$, 
$N_f$-fields $\widetilde{F'}$ are in the fundamental
representation under the $SO(N_c')$ and then 
$F'$ are in the representation $({\bf 1, N_c' })$ 
under the gauge group
while 
$\widetilde{F'}$ are in the representation $({\bf 1, N_c' })$ 
under the gauge group

These additional $2N_f$ $SO(N_c')$ vectors are originating from 
the $SU(N_c)$ chiral mesons $\widetilde{X} Q$ and $X \widetilde{Q}$ respectively.
It is easy to see that from the Figure 5, since the
$D6_{-\omega}$-branes are parallel to the $NS5_{-\omega}$-brane, 
the newly created $N_f$ D4-branes can slide along the plane
consisting of $D6_{-\omega}$-branes and  $NS5_{-\omega}$-brane
arbitrarily(and its mirrors).
Then  strings connecting the $N_f$ $D6_{-\omega}$-branes and $N_c'$
D4-branes will give rise to these additional $2N_f$ $SO(N_c')$ vectors.
 
$\bullet$
$N_f^2$-fields $M$ are in the representation $({\bf 1, 1})$ under the
gauge group

This corresponds to the $SU(N_c)$ chiral meson $Q \widetilde{Q}$ and
the fluctuations of the singlet $M$ correspond to the motion of $N_f$
flavor D4-branes along
(789) directions in Figure 5.

$\bullet$
The $N_c^{'2}$ 
singlet $\Phi$ is in the representation $({\bf 1, adj}) \oplus ({\bf 1,symm
})$ 
under the gauge group  

This corresponds to the $SU(N_c)$ chiral meson $X \widetilde{X}$ and
note that both 
$X$ and $\widetilde{X}$ have representation ${\bf N_c'}$ of $SO(N_c')$.
In general, the fluctuations of the singlet $\Phi$ correspond to the motion of
$N_c'$ D4-branes suspended two $NS5_{\pm \omega}$-branes along the (789)
directions in Figure 5. 

In the dual theory, 
since there exist $N_f$ quarks $q$, $N_f$
quarks $\widetilde{q}$, one bifundamental field $Y$ which will give
rise to the contribution of $N_c'$ and 
its complex conjugate $\widetilde{Y}$ which will give
rise to the contribution of $N_c'$,
the coefficient of the beta function 
of the first gauge group factor with (\ref{num}) is
\bea
b_{SU(\widetilde{N}_c)}^{mag}
= 3\widetilde{N}_c-N_f-N_c'
=2N_f +2N_c'-3N_c
\nonu
\eea
and 
since there exist $2N_f'$ quarks $Q'$, 
one bifundamental field $Y$ which will give
rise to the contribution of $\widetilde{N}_c$, 
its complex conjugate $\widetilde{Y}$ which will give
rise to the contribution of $\widetilde{N}_c$,
$N_f$ fields $F'$, its complex conjugate
$N_f$ fields $\widetilde{F'}$ and the singlet $\Phi$ which will give
rise to $N_c'$, 
the coefficient of the beta function
 is
\bea
b_{SO(N_c')}^{mag} 
= 3(N_c'-2)-2N_f'-2\widetilde{N}_c-2N_f-2N_c' 
=-N_c'+2N_c-4N_f-2N_f'-6.
\nonu
\eea
Therefore, both $SU(\widetilde{N}_c)$ and 
$SO(N_c')$ gauge couplings are IR free
by requiring the negativeness of the coefficients of beta function.
One can rely on the perturbative calculations at low energy 
for this magnetic IR free region $b_{SU(\widetilde{N}_c)}^{mag} < 0$ and 
$b_{SO(N_c')}^{mag} < 0$.
Note that the $SO(N_c')$ fields in the magnetic theory 
are different from those of the electric theory.
Since $b_{SO(N_c')}-b_{SO(N_c')}^{mag} > 0$, $SO(N_c')$ is more
asymptotically free than $SO(N_c')^{mag}$.
Neglecting the $SO(N_c')$ dynamics, the magnetic $SU(\widetilde{N}_c)$
is IR free when $N_f+N_c'< \frac{3}{2} N_c$, as in previous case.

The dual magnetic superpotential, by adding the mass term for $Q$ and
$\widetilde{Q}$
in the electric theory which is equal to put a linear term in $M$ in
the dual magnetic theory, is given by \footnote{There appears a
  mismatch between the number of colors from field theory analysis and
those from brane motion when we take the full dual process
on the two gauge group factors simultaneously \cite{LO}. 
By adding $4N_f'$ D4-branes
to the dual brane configuration without affecting the linking number
counting,
this mismatch can be removed. Similar phenomena occurred in
\cite{BH,Ahn97}.
Then this turned out that there exists a deformation $\Delta W$
generated by the meson $Q' X \widetilde{X} Q'$. This is exactly the
second term, $Q' \Phi Q'$, in (\ref{Wdual}). In previous example,
there is no such deformation term in (\ref{dualW}).  }
\bea
W_{dual} = \left[(\Phi^2 + \cdots) + Q' \Phi Q' +
M q \widetilde{q} + Y \widetilde{F'} \widetilde{q} + \widetilde{Y} q
F' + \Phi Y \widetilde{Y} \right] + m M
\label{Wdual}
\eea
where  the mesons in terms of the fields defined in the electric theory
are
\bea
 M \equiv Q \widetilde{Q}, \qquad \Phi \equiv X \widetilde{X}, \qquad 
F' \equiv \widetilde{X} Q, \qquad \widetilde{F'} \equiv X \widetilde{Q}.
\nonu 
\eea
We abbreviated all the relevant terms and coefficients appearing in
the quartic superpotential for the bifundamentals in electric theory 
(\ref{sup}) and denote them here by $\Phi^2 + \cdots$. 
Here $q$ and $\widetilde{q}$ are fundamental and antifundamental for
the gauge group index respectively and antifundamentals for the flavor index.
Then, $q \widetilde{q}$ has rank $\widetilde{N}_c$ and $m$ has a
rank $N_f$.  Therefore, the F-term condition, the derivative the 
superpotential $W_{dual}$ with respect to $M$, cannot be satisfied 
if the rank $N_f$ exceeds $\widetilde{N}_c$ and the supersymmetry is broken.    
Other F-term equations are satisfied by taking the vacuum expectation 
values of $Y, \widetilde{Y}, F', \widetilde{F'}$ and $Q'$ to vanish.

The classical moduli space of vacua can be obtained from F-term
equations and one gets
\bea
 q  \widetilde{q} +  m & = & 0, \qquad
\widetilde{q} M + F' \widetilde{Y}   =  0, \nonu \\
  M q  + Y \widetilde{F'} & = & 0, \qquad
\widetilde{F'} \widetilde{q} +  \widetilde{Y} \Phi  =  0,
\nonu \\
\widetilde{q} Y & = & 0, \qquad
q F' +  \Phi Y  =  0, \nonu \\
\widetilde{Y} q & = & 0, \qquad
Q' Q' + Y \widetilde{Y}  =  0, 
\nonu \\
\Phi Q' & = & 0.
\nonu
\eea
Then, it is easy to see that there exists a solution
\bea
\widetilde{q} M =0= M q, \qquad
 q  \widetilde{q} +  m  =  0.
\nonu 
\eea
Other F-term equations are satisfied if one takes the zero vacuum
expectation values for the fields $Y, \widetilde{Y}, F', Q'$ and 
$\widetilde{F'}$.
Then the solutions can be written as
\bea
<q >  & = &  \left(
\begin{array}{c}
\sqrt{m} e^{\phi} {\bf 1}_{\widetilde{N}_c}  \\
0
\end{array}
\right),  
< \widetilde{q}> =
 \left(
\begin{array}{cc}
\sqrt{m} e^{-\phi}  {\bf 1}_{\widetilde{N}_c}   &
0
\end{array}
\right), 
<M>  =
 \left(
\begin{array}{cc}
0  & 0 
 \\
0 & \Phi_0  {\bf 1}_{N_f-\widetilde{N}_c} 
\end{array}
\right)
\nonu \\
<Y> & = & <\widetilde{Y}> = <F'> = <\widetilde{F'}> = <Q'>= 0.
\label{vac1}
\eea
Let us expand around a point on (\ref{vac1}), as done in
\cite{ISS}. 
Then the remaining relevant terms of superpotential are given by
\bea
W_{dual}^{rel} & = &  \Phi_0 \left( \delta \varphi  
\; \delta \widetilde{\varphi} + m \right) +
  \delta Z \; \delta \varphi  \; \widetilde{q}_0 
+ \delta \widetilde{Z} \; q_0  
\delta \widetilde{\varphi}
\nonu
\eea
by following the similar fluctuations 
for the various fields as in \cite{Ahn07}.
Note that there exist also four kinds of terms, 
the vacuum  $<q>$ multiplied by 
$\delta \widetilde{Y} \delta F'$,  
the vacuum  $<\widetilde{q}>$ multiplied by $\delta \widetilde{F'} 
\delta Y$, 
the vacuum  $<\Phi>$ multiplied by $\delta Y 
\delta \widetilde{Y}$, and 
the vacuum  $<\Phi>$ multiplied by $\delta Q' 
\delta Q'$. However,
by redefining these, they do not enter the 
contributions for the one loop result, up to quadratic order. 
As done in \cite{Shih}, one gets 
that $m_{\Phi_0}^2$ will contain $(\log 4 -1) > 0$ implying that these
are stable.

\section{Nonsupersymmetric meta-stable brane configuration
of $SU(N_c) \times SO(N_c')$ gauge theory }

Since the electric superpotential (\ref{sup}) 
vanishes for $\theta=0$ and $\omega=\frac{\pi}{2}$,
the corresponding magnetic superpotential in (\ref{Wdual}) does not 
contain  the terms $\Phi^2 + \cdots$ and it becomes
\bea
W_{dual} = \left(Q' \Phi Q' +
M q \widetilde{q} + Y \widetilde{F'} \widetilde{q} + \widetilde{Y} q
F' + \Phi Y \widetilde{Y} \right) + m M.
\nonu
\eea
Now we recombine $\widetilde{N}_c$ D4-branes among $N_f$ flavor D4-branes 
connecting between $D6_{\omega=\frac{\pi}{2}}=D6$-branes and 
$NS5_{\omega=\frac{\pi}{2}}=NS5_R'$-brane with those 
connecting between $NS5_R'$-brane and $NS5_{-\theta=0}=NS5_R$-brane(and
its mirrors) 
and push
them in $+v$ direction from Figure 5. 
Of course their mirrors will move
to $-v$ direction in a ${\bf Z}_2$ symmetric manner due to the 
$O6^{+}$-plane. 
After this procedure, there are no color D4-branes between 
$NS5_R'$-brane and $NS5_R$-brane.
For the flavor D4-branes, we are left with only 
$(N_f-\widetilde{N}_c)$ D4-branes(and its mirrors).  

Then the minimal energy supersymmetry breaking brane configuration is
shown in Figure 6.
If we ignore all the branes except $NS5_R'$-brane, $NS5_R$-brane, 
D6-branes, $(N_f-\widetilde{N}_c)$ D4-branes and 
$\widetilde{N}_c$ D4-branes, 
as observed already, 
then this brane configuration 
corresponds to  the minimal energy supersymmetry breaking brane
configuration
for the ${\cal N}=1$ SQCD with the magnetic gauge group 
$SU(\widetilde{N}_c)$ with
$N_f$ massive flavors \cite{OO,FGU,BGHSS}. 
Note that $N_c'$ D4-branes can slide $w$
direction for this brane configuration.

\begin{figure}[ht]
   \epsfxsize=5in 
\centerline{\epsffile{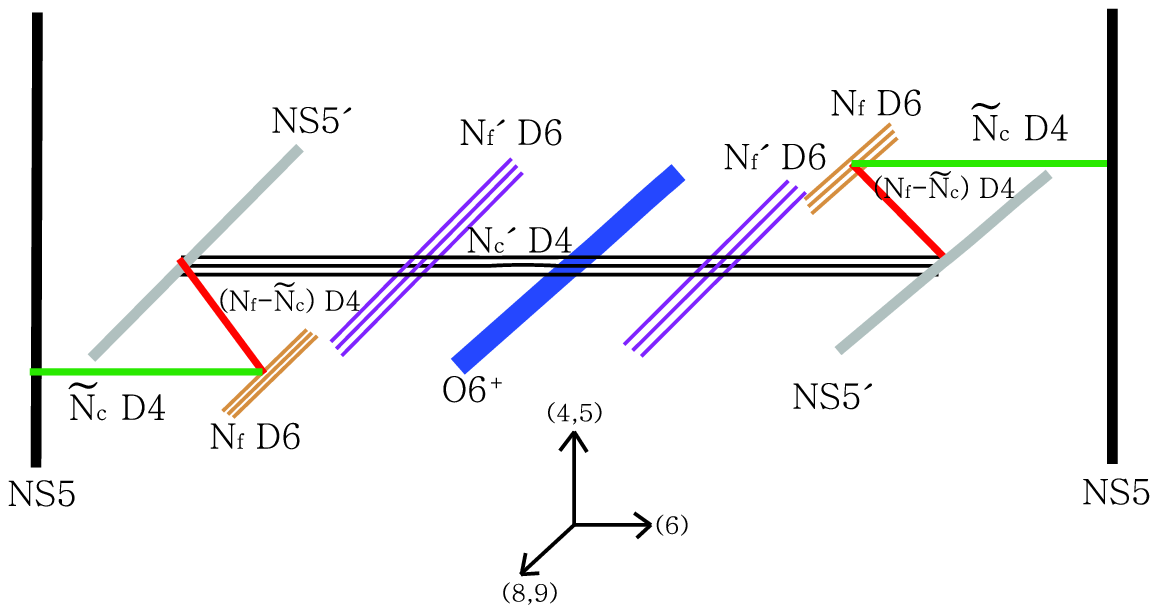}}
   \caption[FIG. \arabic{figure}.]{ 
The nonsupersymmetric minimal energy brane configuration of 
$SU(\widetilde{N}_c=N_f +N_c'-N_c) \times SO(N_c')$ 
with $N_f$ chiral multiplets $q$,
$N_f$ chiral multiplets $\widetilde{q}$, $2N_f'$ chiral multiplets
$Q'$, 
the flavor singlet
bifundamental field $Y$ and its 
complex conjugate bifundamental field $\widetilde{Y}$ and gauge singlets.
The $N_c'$ D4-branes and $2(N_f-\widetilde{N}_c)$ D4-branes can slide
$w$ direction freely in a ${\bf Z}_2$ symmetric way.  }
\label{fig6}
\end{figure}

The type IIA/M-theory brane construction for the ${\cal N}=2$
gauge theory was described by \cite{LL} and after lifting 
the type IIA description we explained so far to M-theory, the 
corresponding magnetic M5-brane configuration with equal mass for the
quarks where the gauge group is given by 
$SU(\widetilde{N}_c) \times SO(N_c')$, in a background space of $x t = 
(-1)^{N_f+N_f'} v^{2N_f'+4} 
\prod_{k=1}^{N_f} (v^2 -e_k^2)$ where this four dimensional space
replaces (45610) directions,
is characterized by 
\bea
t^4 + ( v^{\widetilde{N}_c }  + \cdots ) t^3 + 
( v^{N_c'} + \cdots) t^2 +  (v^{\widetilde{N}_c}  + \cdots ) 
t + v^{2N_f'+4}
\prod_{k=1}^{N_f} (v^2 -e_k^2 ) =0.
\nonu
\eea

From this curve 
of quartic equation for $t$ above, the asymptotic regions 
can be classified by looking at 
the first two terms providing $NS5_R$-brane asymptotic region, 
next two terms providing $NS5_R'$-brane asymptotic region,
next two terms providing $NS5_L'$-brane asymptotic region,
 and 
the final two terms giving $NS5_L$-brane asymptotic region
as follows:

1. $v \rightarrow \infty$ limit implies
\bea
w & \rightarrow & 0, \quad y \sim    v^{\widetilde{N}_c} + \cdots \quad
\mbox{$NS5_{R}$ 
asymptotic region}, \nonu \\
w & \rightarrow  & 0, \quad y \sim    
v^{2N_f+2N_f'-\widetilde{N}_c+4} + \cdots \quad
\mbox{$NS5_{L}$ asymptotic region}.   
\nonu
\eea

2.  $w \rightarrow \infty$ limit implies
\bea
v & \rightarrow &   -m, \quad 
y \sim  w^{\widetilde{N}_c-N_c'}
 +\cdots
\quad \mbox{$NS5_{L}'$ asymptotic region}, 
\nonu
\\
v & \rightarrow &  +m, \quad  
y \sim w^{N_c'-\widetilde{N}_c}
+\cdots
\quad \mbox{$NS5_{R}'$ asymptotic region}. 
\nonu
\eea
Now the two $NS5_{L,R}'$-branes are moving in the
$\pm v$ direction holding everything else fixed instead of moving
D6-branes in the $\pm v$ direction. Then the mirrors of D4-branes are
moved appropriately. 
The harmonic function sourced by the D6-branes can be written
explicitly by summing of three contributions from the 
$N_f$ and $N_f'$ D6-branes(and its mirrors) plus an O6-plane, 
and similar analysis to solve
the differential equation and find out the nonholomorphic curve can be
done \cite{BGHSS,Ahn06-1,Ahn07,Ahn07-1,Ahn07-2}. 
In this case also, we expect an instability from a new M5-brane mode.

\section{Discussions}

So far, we have dualized only the first gauge group factor in the
gauge group $SU(N_c) \times SO(N_c')$. What happens if we dualize the
second gauge group factor $SO(N_c')$?(For the case $SU(N_c) \times
SU(N_c')$, the behavior of dual for the second gauge group will be the
same as when we take the dual for the first gauge group factor.)
This can be done by moving the $NS5_{\theta}$-brane and
$N_f'$ $D6_{\theta}$-branes that can be located at the nonzero $v$
coordinate for massive quarks $Q'$, 
to the right passing through O6-plane(and their mirrors to the left).
According to the linking number counting, one obtains the dual gauge
group
$SU(N_c) \times SO(\widetilde{N}_c'=2N_c+2N_f'-N_c'+4)$.
One can easily see that there is a creation of $N_f'$ D4-branes
connecting $NS5_{\theta}$-brane and $D6_{\theta}$-branes(and its mirrors). 
Then from the brane configuration, there exist the additional
$2N_f'$ $SU(N_c)$
quarks  originating from the $SO(N_c')$ chiral mesons $Q'X \equiv
\widetilde{F'}$ 
and $Q'\widetilde{X} \equiv F'$. The deformed superpotential $\Delta
W=Q' X \widetilde{X} Q'$
can be interpreted as the mass term of $ F' \widetilde{F'}$.    
Then one can write dual magnetic superpotential in this case. 
However, it is not clear how the recombination of color and flavor 
D4-branes and splitting procedure between them
in the construction of meta-stable vacua arises since there is no
extra NS5-brane between two $NS5_{\pm \theta}$-branes. If there exists
an extra NS5-brane at the origin of our brane configuration(then the
gauge group and matter contents will change), it would
be possible to construct the corresponding meta-stable brane
configuration. It would be interesting to study these more in the future.  

As already mentioned in \cite{Ahn07-1} and section 4, 
the matter contents in \cite{ILS} are different from the ones in
section 4 with the same gauge group. In other words, 
 the theory of $SU(N_c) \times SO(N_c')$ with $X$, which
transform as fundamental in $SU(N_c)$ and vector in $SO(N_c')$, a
antisymmetric tensor $A$ in $SU(N_c)$, as well as
fundamentals for $SU(N_c)$ and vectors for $SO(N_c')$ 
can confine either $SU(N_c)$ factor or $SO(N_c')$ factor. 
This theory can be described by the web of branes in the presence of 
$O4^{-}$-plane and orbifold fixed points. 
With two NS5-branes and $O4^{-}$-plane,
by modding out ${\bf Z}_3$ symmetry acting on $(v,w)$ as
$(v,w) \rightarrow (v \exp(\frac{2\pi i}{3}), w \exp(\frac{2\pi
  i}{3}))$,
the resulting gauge group will be $SU(N_c) \times SO(N_c+4)$ with above
matter contents \cite{LPT}.
Similar analysis for $SU(N_c) \times Sp(\frac{N_c}{2}-2)$ gauge group
with opposite $O4^{+}$-plane can be done.
Then in this case, the matter in $SU(N_c)$ will be 
a symmetric tensor $S$ and
other matter contents are present also.
It would be interesting to see whether this gauge theory and
corresponding brane configuration will provide a meta-stable vacuum.

Let us comment on other possibility
where
the gauge group is given by $SU(N_c) \times Sp(N_c')$ and the matter 
contents are given by

$\bullet$
$N_f$ chiral multiplets $Q$ are 
in the fundamental representation under the $SU(N_c)$,
$N_f$ chiral multiplets $\widetilde{Q}$ are in the antifundamental
representation under the $SU(N_c)$ and then $Q$ are in the
representation $({\bf N_c,1
})$ while $\widetilde{Q}$ are in the representation $({\bf \overline{N_c}, 1})$
under the gauge group

$\bullet$
$2N_f'$ chiral multiplets $Q'$ are 
in the fundamental representation under the $Sp(N_c')$ and then $Q'$ are in the
representation $({\bf 1, 2N_c'
})$
under the gauge group

$\bullet$
The flavor singlet field $X$ is 
in the bifundamental representation $({\bf N_c, 2N_c' })$ 
under the gauge group and the flavor singlet $\widetilde{X}$
 is 
in the bifundamental representation $({\bf \overline{N_c}, 2N_c'})$ 
under the gauge group

One can compute the coefficients of beta functions of the each gauge
group factor, as we did for previous examples.

The type IIA brane configuration of an electric theory is exactly the
same as the Figure 4 except the RR charge O6-plane with negative sign.  
The classical superpotential \footnote{The superpotential for the $Sp(N_c')$
sector is given by $W=X \phi_A \widetilde{X} + X \phi_S \widetilde{X}+
\tan \theta \tr \phi_S^2 -\frac{1}{\tan \theta} \tr \phi_A^2$ where 
$\phi_S$ and $\phi_A$ are an adjoint 
field(symmetric tensor) and an antisymmetric tensor for
  $Sp(N_c')$ \cite{CSST}. Note that there is a sign change in the
  second trace term of the 
superpotential in (\ref{sup1}), compared to (\ref{sup}).}
is
given by \cite{LO}
\bea
W= -\frac{1}{4} \left[ \frac{1}{4 \tan(\omega - \theta)} +
  \frac{1}{\tan 2\theta} \right]
\tr (X \widetilde{X})^2 - \frac{ \tr X \widetilde{X}
\widetilde{X} X }{4 \sin 2\theta }
+ \frac{(\tr
X \widetilde{X})^2}{4 N_c \tan(\omega-\theta)}.
\label{sup1}
\eea
In this case, when $\theta$ approaches $\frac{\pi}{2}$ and
$\omega$
approaches $0$, 
then this superpotential vanishes.  

The dual magnetic gauge group
is given by $SU(\widetilde{N}_c=N_f+2N_c'-N_c) \times Sp(N_c')$ with
the same number of colors of dual theory as those in previous cases 
and the matter contents are given by

$\bullet$ 
$N_f$ chiral multiplets $q$ are 
in the fundamental representation under the $SU(\widetilde{N}_c)$,
$N_f$ chiral multiplets $\widetilde{q}$ are in the antifundamental
representation under the $SU(\widetilde{N}_c)$ and then $q$ are in the
representation $({\bf \widetilde{N}_c,1
})$ while $\widetilde{q}$ are in the representation 
$({\bf \overline{\widetilde{N}_c}, 1})$
under the gauge group

$\bullet$
$2N_f'$ chiral multiplets $Q'$ are 
in the fundamental representation under the $Sp(N_c')$ and then $Q'$ are in the
representation $({\bf 1, 2N_c'
})$
under the gauge group

$\bullet$
The flavor singlet field $Y$ is 
in the bifundamental representation $({\bf \widetilde{N}_c, 2N_c' })$ 
under the gauge group and its complex conjugate field $\widetilde{Y}$
 is 
in the bifundamental representation $({\bf \overline{\widetilde{N}_c}, 2N_c'})$ 
under the gauge group

There are $(N_f+2N_c')^2$ 
gauge singlets in the first dual gauge group factor

$\bullet$
$N_f$-fields $F'$ are in the fundamental representation under the
$Sp(N_c')$, 
$N_f$-fields $\widetilde{F'}$ are in the fundamental
representation under the $Sp(N_c')$ and then 
$F'$ are in the representation $({\bf 1, 2N_c' })$ 
under the gauge group
while 
$\widetilde{F'}$ are in the representation $({\bf 1, 2N_c' })$ 
under the gauge group

$\bullet$
$N_f^2$-fields $M$ are in the representation $({\bf 1, 1})$ under the
gauge group

$\bullet$
The $4N_c^{'2}$ 
singlet $\Phi$ is in the representation $({\bf 1, adj}) \oplus ({\bf 1,antisymm
})$ 
under the gauge group  

The dual magnetic superpotential for arbitrary angles is given by (\ref{Wdual}) with
appropriate $Sp(N_c')$ invariant metric $J$. The stability analysis
can be done similarly.

After following the procedure from Figure 4 to Figure 5 with opposite
RR charge for O6-plane and by taking the limit where $\theta
\rightarrow \frac{\pi}{2}$ and $\omega \rightarrow 0$,  
the minimal energy supersymmetry breaking brane configuration is
shown in Figure 7.

\begin{figure}[ht]
   \epsfxsize=5in 
\centerline{\epsffile{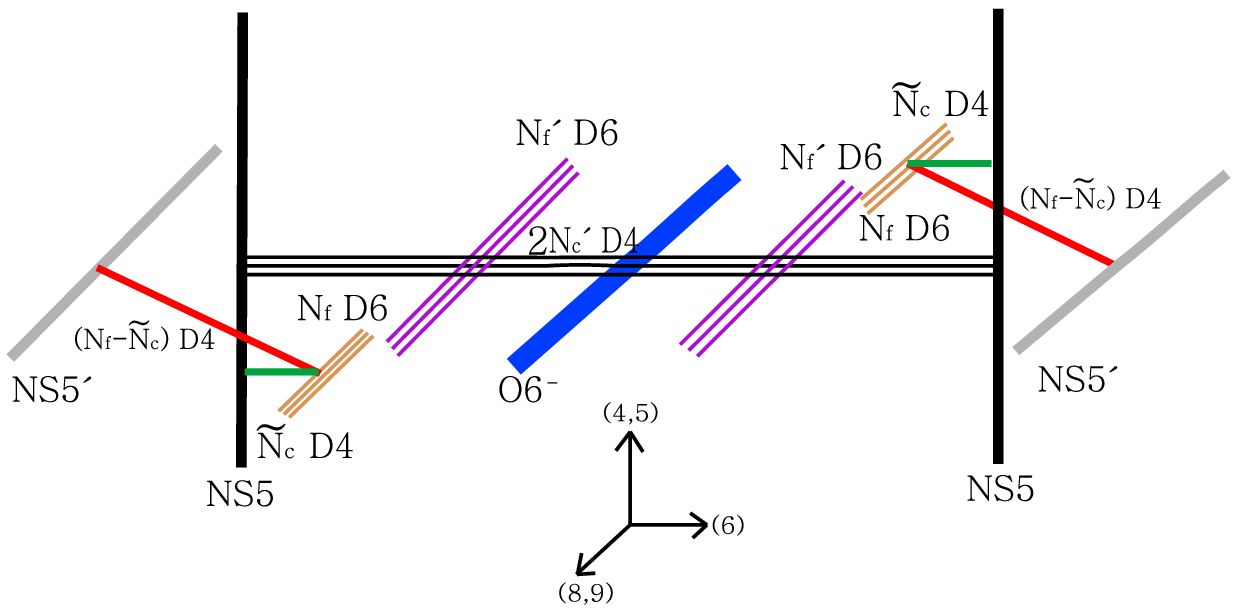}}
   \caption[FIG. \arabic{figure}.]{ 
The nonsupersymmetric minimal energy brane configuration of 
$SU(\widetilde{N}_c=N_f +2N_c'-N_c) \times Sp(N_c')$ 
with $N_f$ chiral multiplets $q$,
$N_f$ chiral multiplets $\widetilde{q}$, $2N_f'$ chiral multiplets
$Q'$, 
the flavor singlet
bifundamental field $Y$ and its 
complex conjugate bifundamental field $\widetilde{Y}$ and gauge singlets.
Note the RR charge of O6-plane is negative and its charge is
equivalent to $-4$ D6-branes. The $2N_c'$ D4-branes and 
$2(N_f-\widetilde{N}_c)$ D4-branes can slide
$w$ direction freely in a ${\bf Z}_2$ symmetric way. }
\label{fig7}
\end{figure}

Compared to the previous nonsupersymmetric brane configuration in
Figure 6, the role of NS5-brane and NS5'-brane is interchanged to each
other: undoing the Seiberg dual in the context of \cite{FGU}. 
This kind of  
feature of recombination and splitting between color D4-branes and
flavor D4-branes occurs in \cite{Ahn07-1}.
At the electric brane configuration, 
$N_f$ D6-branes are perpendicular to NS5-brane
and this leads to the coupling between the quarks and adjoint in the
superpotential.
However, the overall coefficient function including this extra terms
vanishes and eventually the whole electric superpotential will vanish
according to the above limit we take.

From the quartic equation with the presence of opposite
RR charge for O6-plane, in a background space of $x t = 
(-1)^{N_f+N_f'} v^{2N_f'-4} 
\prod_{k=1}^{N_f} (v^2 -e_k^2)$,
\bea
t^4 + ( v^{\widetilde{N}_c }  + \cdots ) t^3 + 
( v^{N_c'} + \cdots) t^2 +  (v^{\widetilde{N}_c}  + \cdots ) 
t + v^{2N_f'-4}
\prod_{k=1}^{N_f} (v^2 -e_k^2 ) =0,
\nonu
\eea
the asymptotic regions 
can be classified as follows:

1. $v \rightarrow \infty$ limit implies
\bea
w & \rightarrow & 0, \quad y \sim     v^{N_c'-\widetilde{N}_c} \cdots \quad
\mbox{$NS5_{R}$ 
asymptotic region}, \nonu \\
w & \rightarrow  & 0, \quad y \sim    v^{\widetilde{N}_c-N_c'}  + \cdots \quad
\mbox{$NS5_{L}$ asymptotic region}.   
\nonu
\eea

2.  $w \rightarrow \infty$ limit implies
\bea
v & \rightarrow &   -m, \quad 
y \sim   w^{2N_f+2N_f'-\widetilde{N}_c-4}
 +\cdots
\quad \mbox{$NS5_{L}'$ asymptotic region}, 
\nonu
\\
v & \rightarrow &  +m, \quad  
y \sim w^{\widetilde{N}_c}
+\cdots
\quad \mbox{$NS5_{R}'$ asymptotic region}. 
\nonu
\eea

In \cite{IT}, the $SU(7) \times \widetilde{Sp}(1)$ model  and 
$SU(9) \times \widetilde{Sp}(2)$ model  can be
obtained by dualizing the $SU(7) \times SU(2)$ model with 
a bifundamental and two
antifundamentals for $SU(7)$ and a fundamental for $SU(2)$
and the $SU(9) \times SU(2)$
with
a bifundamental and two
antifundamentals for $SU(9)$ and a fundamental for $Sp(1)$
respectively(Note that $Sp(1) \sim SU(2)$).
The matter contents in an electric theory are different from those in
previous paragraph.
The matter contents in the magnetic description are given by 
an antisymmetric tensor and a fundamental in the first gauge group
as well as a bifundamental, a fundamental in the second gauge group
and two antifundamentals in the first gauge group.
There exists a nonzero dual magnetic superpotential. 
Also the dual description  the $SU(7) \times \widetilde{Sp}(1)$ model  and 
$SU(9) \times \widetilde{Sp}(2)$ model can be constructed from 
the antisymmetric models of Affleck-Dine-Seiberg by 
gauging a maximal flavor symmetry
and adding the extra matter to cancel all anomalies and extra flavor. 

On the other hand, 
the models $SU(2N_c+1) \times SU(2)$ have its brane box model
description in \cite{HZ} where the above examples correspond to $N_c=3$
and $N_c=4$ respectively. 
In particular, the case where $N_c=1$(the gauge group is 
$SU(3) \times SU(2)$, i.e., $(3,2)$ model \cite{ADS}) 
was described by brane box model
with superpotential or without superpotential. 
Then it would be interesting to obtain the
Seiberg dual for these models using brane box model and look for the
possibility of having meta-stable vacua for these models.  
Moreover, this gauge theory was generalized to $SU(2N_c+1) \times Sp(N_c')$
model with 
a bifundamental and $2N_c'$
antifundamentals for $SU(2N_c+1)$ and a fundamental for $Sp(N_c')$
and its dual description $SU(2N_c+1) 
\times Sp(\widetilde{N}_c'=N_c-N_c'-1)$ with
a bifundamental and $2N_c'$
antifundamentals for $SU(2N_c+1)$ and a fundamental for $Sp(N_c')$ 
as well as two gauge singlets \cite{IT}.
For the particular range of $N_c$, the dual theory is IR free, not
asymptotically free.

According to \cite{Berkooz}, $SU(2N_c)$ with antisymmetric tensor and
antifundamentals can be described by two gauge groups $Sp(2N_c-4) \times
SU(2N_c)$
with bifundamental and antifundamentals for $SU(2N_c)$. Some of the
brane 
realization with zero superpotential was given in the brane box model
in \cite{HZ}. Similarly from
the result of \cite{Pouliot} by following the method of
\cite{Berkooz}, 
the dual description for 
$SU(2N_c+1)$ with antisymmetric tensor and fundamentals can be
represented by two gauge group factors. This dual theory breaks the
supersymmetry at the tree level.  Similar discussions are present in \cite{PS}.
Then it would be interesting to construct the corresponding Seigerg
dual and see how the electric theory and its magnetic theory can be
mapped into each other in the brane box model. 

Ther are also different directions concerning on the meta-stable vacua
in different contexts and some of the relevant works are present in
\cite{Murthy}-\cite{ABFK} where some of them use anti D-branes 
and some of them describe the type IIB theory 
and it would be interesting to find out
how similarities if any appear and what are the differences in what sense 
between the present work and those works. 

\vspace{.7cm}
\centerline{\bf Acknowledgments}

I would like to thank A. Hanany and K. Landsteiner for discussions. 
This work was supported by grant No.
R01-2006-000-10965-0 from the Basic Research Program of the Korea
Science \& Engineering Foundation.


\begin{thebibliography}{99}

\bibitem{ISS}
  K.~Intriligator, N.~Seiberg and D.~Shih,
  ``Dynamical SUSY breaking in meta-stable vacua,''
  JHEP {\bf 0604}, 021 (2006)
  [arXiv:hep-th/0602239].

\bibitem{IS}
  K.~Intriligator and N.~Seiberg,
  ``Lectures on supersymmetry breaking,''
  [arXiv:hep-ph/0702069].

\bibitem{GK}
  A.~Giveon and D.~Kutasov,
  ``Brane dynamics and gauge theory,''
  Rev.\ Mod.\ Phys.\  {\bf 71}, 983 (1999)
  [arXiv:hep-th/9802067].

\bibitem{ILS}
  K.~A.~Intriligator, R.~G.~Leigh and M.~J.~Strassler,
  ``New examples of duality in chiral and nonchiral supersymmetric gauge
  theories,''
  Nucl.\ Phys.\ B {\bf 456}, 567 (1995)
  [arXiv:hep-th/9506148].

\bibitem{BH}
  J.~H.~Brodie and A.~Hanany,
  ``Type IIA superstrings, chiral symmetry, and N = 1 4D gauge theory
  dualities,''
  Nucl.\ Phys.\ B {\bf 506}, 157 (1997)
  [arXiv:hep-th/9704043].

\bibitem{BIWW}
  E.~Barnes, K.~Intriligator, B.~Wecht and J.~Wright,
  ``N = 1 RG flows, product groups, and a-maximization,''
  Nucl.\ Phys.\  B {\bf 716}, 33 (2005)
  [arXiv:hep-th/0502049].

\bibitem{Ahn07-2}
  C.~Ahn,
  ``Meta-Stable Brane Configuration and Gauged Flavor Symmetry,''
  [arXiv:hep-th/0703015].

\bibitem{Ahn07-1}
  C.~Ahn,
  ``More on meta-stable brane configuration,''
  [arXiv:hep-th/0702038].

\bibitem{Ahn07}
  C.~Ahn,
  ``Meta-stable brane configuration with orientifold 6 plane,''
  [arXiv:hep-th/0701145], to appear in JHEP.

\bibitem{Ahn06-1}
  C.~Ahn,
  ``M-theory lift of meta-stable brane configuration in symplectic and
  orthogonal gauge groups,''
  Phys.\ Lett.\  B {\bf 647}, 493 (2007)
  [arXiv:hep-th/0610025].

\bibitem{Ahn06}
  C.~Ahn,
  ``Brane configurations for nonsupersymmetric meta-stable vacua in SQCD with
  adjoint matter,''
  Class.\ Quant.\ Grav.\  {\bf 24}, 1359 (2007)
  [arXiv:hep-th/0608160].

\bibitem{OO}
  H.~Ooguri and Y.~Ookouchi,
  ``Meta-stable supersymmetry breaking vacua on intersecting branes,''
  Phys.\ Lett.\ B {\bf 641}, 323 (2006)
  [arXiv:hep-th/0607183].

\bibitem{FGU}
  S.~Franco, I.~Garcia-Etxebarria and A.~M.~Uranga,
  ``Non-supersymmetric meta-stable vacua from brane configurations,''
  JHEP {\bf 0701}, 085 (2007)
  [arXiv:hep-th/0607218].

\bibitem{BGHSS}
  I.~Bena, E.~Gorbatov, S.~Hellerman, N.~Seiberg and D.~Shih,
  ``A note on (meta)stable brane configurations in MQCD,''
  JHEP {\bf 0611}, 088 (2006)
  [arXiv:hep-th/0608157].

\bibitem{LO}
  E.~Lopez and B.~Ormsby,
  ``Duality for SU x SO and SU x Sp via branes,''
  JHEP {\bf 9811}, 020 (1998)
  [arXiv:hep-th/9808125].

\bibitem{HW}
  A.~Hanany and E.~Witten,
  ``Type IIB superstrings, BPS monopoles, and three-dimensional gauge
  dynamics,''
  Nucl.\ Phys.\ B {\bf 492}, 152 (1997)
  [arXiv:hep-th/9611230].

\bibitem{Shih}
  D.~Shih,
  ``Spontaneous R-Symmetry Breaking in O'Raifeartaigh Models,''
  [arXiv:hep-th/0703196].

\bibitem{Witten}
  E.~Witten,
  ``Solutions of four-dimensional field theories via M-theory,''
  Nucl.\ Phys.\  B {\bf 500}, 3 (1997)
  [arXiv:hep-th/9703166].

\bibitem{LL}
  K.~Landsteiner and E.~Lopez,
  ``New curves from branes,''
  Nucl.\ Phys.\  B {\bf 516}, 273 (1998)
  [arXiv:hep-th/9708118].

\bibitem{LLL}
  K.~Landsteiner, E.~Lopez and D.~A.~Lowe,
  ``Supersymmetric gauge theories from branes and orientifold six-planes,''
  JHEP {\bf 9807}, 011 (1998)
  [arXiv:hep-th/9805158].

\bibitem{AOT}
  C.~Ahn, K.~Oh and R.~Tatar,
  ``Comments on SO/Sp gauge theories from brane configurations with an  O(6)
  plane,''
  Phys.\ Rev.\ D {\bf 59}, 046001 (1999)
  [arXiv:hep-th/9803197].

\bibitem{LLL1}
  K.~Landsteiner, E.~Lopez and D.~A.~Lowe,
  ``Duality of chiral N = 1 supersymmetric gauge theories via branes,''
  JHEP {\bf 9802}, 007 (1998)
  [arXiv:hep-th/9801002].

\bibitem{BHKL}
  I.~Brunner, A.~Hanany, A.~Karch and D.~Lust,
  ``Brane dynamics and chiral non-chiral transitions,''
  Nucl.\ Phys.\ B {\bf 528}, 197 (1998)
  [arXiv:hep-th/9801017].

\bibitem{EGKT}
  S.~Elitzur, A.~Giveon, D.~Kutasov and D.~Tsabar,
  ``Branes, orientifolds and chiral gauge theories,''
  Nucl.\ Phys.\ B {\bf 524}, 251 (1998)
  [arXiv:hep-th/9801020].

\bibitem{CSST}
  C.~Csaki, M.~Schmaltz, W.~Skiba and J.~Terning,
  ``Gauge theories with tensors from branes and orientifolds,''
  Phys.\ Rev.\ D {\bf 57}, 7546 (1998)
  [arXiv:hep-th/9801207].

\bibitem{Ahn97}
  C.~Ahn, K.~Oh and R.~Tatar,
  ``Branes, geometry and N = 1 duality with product gauge groups 
of SO and
  Sp,''
  J.\ Geom.\ Phys.\  {\bf 31}, 301 (1999)
  [arXiv:hep-th/9707027].

\bibitem{LPT}
  J.~D.~Lykken, E.~Poppitz and S.~P.~Trivedi,
  ``M(ore) on chiral gauge theories from D-branes,''
  Nucl.\ Phys.\  B {\bf 520}, 51 (1998)
  [arXiv:hep-th/9712193].

\bibitem{IT}
  K.~A.~Intriligator and S.~D.~Thomas,
  ``Dual descriptions of supersymmetry breaking,''
  [arXiv:hep-th/9608046].

\bibitem{HZ}
  A.~Hanany and A.~Zaffaroni,
  ``On the realization of chiral four-dimensional gauge theories using
  branes,''
  JHEP {\bf 9805}, 001 (1998)
  [arXiv:hep-th/9801134].

\bibitem{ADS}
  I.~Affleck, M.~Dine and N.~Seiberg,
   ``Dynamical Supersymmetry Breaking In Four-Dimensions And Its
  Phenomenological Implications,''
  Nucl.\ Phys.\  B {\bf 256}, 557 (1985).

\bibitem{Berkooz}
  M.~Berkooz,
  ``The Dual of supersymmetric SU(2k) with an antisymmetric tensor and
  composite dualities,''
  Nucl.\ Phys.\  B {\bf 452}, 513 (1995)
  [arXiv:hep-th/9505067].

\bibitem{Pouliot}
  P.~Pouliot,
  ``Duality in SUSY $SU(N)$ with an Antisymmetric Tensor,''
  Phys.\ Lett.\  B {\bf 367}, 151 (1996)
  [arXiv:hep-th/9510148].

\bibitem{PS}
  P.~Pouliot and M.~J.~Strassler,
  ``Duality and Dynamical Supersymmetry Breaking in $Spin(10)$ with a Spinor,''
  Phys.\ Lett.\  B {\bf 375}, 175 (1996)
  [arXiv:hep-th/9602031].

\bibitem{Murthy}
  S.~Murthy,
  ``On supersymmetry breaking in string theory from gauge theory in a throat,''
  [arXiv:hep-th/0703237].

\bibitem{Argurio:2007qk}
  R.~Argurio, M.~Bertolini, S.~Franco and S.~Kachru,
  ``Metastable vacua and D-branes at the conifold,''
  [arXiv:hep-th/0703236].

\bibitem{Giveon:2007fk}
  A.~Giveon and D.~Kutasov,
  ``Gauge symmetry and supersymmetry breaking from intersecting branes,''
  [arXiv:hep-th/0703135].

\bibitem{Antebi:2007xw}
  Y.~E.~Antebi and T.~Volansky,
  ``Dynamical supersymmetry breaking from simple quivers,''
  [arXiv:hep-th/0703112].

\bibitem{Wijnholt:2007vn}
  M.~Wijnholt,
  ``Geometry of particle physics,''
  [arXiv:hep-th/0703047].

\bibitem{Heckman:2007wk}
  J.~J.~Heckman, J.~Seo and C.~Vafa,
  ``Phase structure of a brane/anti-brane system at large N,''
  [arXiv:hep-th/0702077].

\bibitem{Tatar:2006dm}
  R.~Tatar and B.~Wetenhall,
  ``Metastable vacua, geometrical engineering and MQCD transitions,''
  JHEP {\bf 0702}, 020 (2007)
  [arXiv:hep-th/0611303].

\bibitem{Verlinde:2006bc}
  H.~Verlinde,
  ``On metastable branes and a new type of magnetic monopole,''
  [arXiv:hep-th/0611069].

\bibitem{Aganagic:2006ex}
  M.~Aganagic, C.~Beem, J.~Seo and C.~Vafa,
  ``Geometrically induced metastability and holography,''
  [arXiv:hep-th/0610249].

\bibitem{ABFK}
  R.~Argurio, M.~Bertolini, S.~Franco and S.~Kachru,
  ``Gauge / gravity duality and meta-stable dynamical supersymmetry breaking,''
  JHEP {\bf 0701}, 083 (2007)
  [arXiv:hep-th/0610212].



\end{thebibliography}
\end{document}